\documentclass[twocolumn,prl,sort&compress,floatfix]{revtex4-1}
\usepackage{amssymb}
\usepackage{amsmath}
\usepackage{mathrsfs}
\usepackage{bm}
\usepackage[dvipsnames]{xcolor}
\usepackage[sort&compress]{natbib}

\usepackage{graphicx}

\usepackage{color}
\definecolor{red}{rgb}{1,0,0}
\definecolor{blue}{rgb}{0,0,1}

\newcommand{\red}[1]{\textcolor{black}{#1}}

\begin{document}
\title{Ring Polymers: Threadings, Knot Electrophoresis and Topological Glasses}

\author{Davide Michieletto$^{1,\dagger}$, Davide Marenduzzo $^{1}$ Enzo Orlandini$^{2}$ and Matthew S. Turner$^{3}$}

\affiliation{%
$^{1}$SUPA, School of Physics and Astronomy, University of Edinburgh, Peter Guthrie Tait Road, Edinburgh, EH9 3FD, UK; \\ 
$^{2}$Dipartimento di Fisica e Astronomia, Sezione INFN, Università di Padova, Via Marzolo 8, 35131 Padova, Italy;\\
$^{3}$Department of Physics and Centre for Complexity Science, University of Warwick, Coventry CV4 7AL, UK;\\
$^\dagger$davide.michieletto@ed.ac.uk\\
}

\begin{abstract}
Elucidating the physics of a concentrated suspension of ring polymers, or of an ensemble of ring polymers in a complex environment, is an important outstanding question in polymer physics. Many of the characteristic features of these systems arise due to topological interactions between polymers, or between the polymers and the environment, and it is often challenging to describe this quantitatively. Here we review recent research which suggests that a key role is played by inter-ring threadings (or penetrations), which become more abundant as the ring size increases. As we discuss, the physical consequences of such threadings are far-reaching: for instance, they lead to a topologically-driven glassy behaviour of ring polymer melts under pinning perturbations, while they can also account for the shape of experimentally observed patterns in two-dimensional gel electrophoresis of DNA knots.
\end{abstract}

\maketitle

\section{Introduction} 

Topology and entanglement often play a major and sometimes puzzling role in physical systems. Scientific interest in these topics date back to Lord Kelvin's theory of knotted ether tubes, followed by mathematical work on knot theory by Tait and Maxwell~\cite{knott2015life,Thomson1867,Thomson1869,Maxwell1890}. In a visionary poem, Maxwell also makes an intriguing link between knots and liquid vortices~\cite{Maxwellpoetry}. Whilst ether theory was later shown to be incorrect, the connection to liquid vortices was physically sound, and, more than a century after the creation of knot theory, knotted fields have been generated in the lab with light~\cite{Kedia2013}, in fluids~\cite{Kleckner2013} and in liquid crystals~\cite{Machon2013}.

In this work, we shall focus on a particular example of entanglement which arises in dense solutions of ring polymers. The concept of entanglement originates in systems of fluctuating filaments such as polymers, although it can also be defined for defect lines in liquid crystal, as well as for optical or fluid vortices. Entanglements, (understood as restrictions in the configurational space or dynamical relaxation due to uncrossability constraints) can occur within the same filament (self-entanglement), between filaments (mutual entanglement), or between filaments and obstacles in the environment. 
If the filaments are closed to form fluctuating rings, these entanglements can often get trapped into knots or links that can be identified mathematically through {\it topological invariants} -- i.e., quantities which only depend on the global filament topology (and not, for instance, on other details of their instantaneous configurations).
Once a filament is knotted, two (or more) filaments are linked, or a filament is linked to an external obstacle (e.g., a long rod), the knot or link type can only be changed if strand passage within or between filaments is allowed. 
A biologically relevant example is given by the action of specific enzymes, type II topoisomerases, that bind and cut regions of circular double-stranded (ds) DNA to simplify the topological self-entanglements (knots) or mutual entanglement (catenanes) that may develop, for instance, during DNA replication.
A striking example of topological mutual entanglement is instead observed in the DNA kinetoplast where  thousands of interlinked DNA minicircles form a spanning network filling the mithocondria whose topology during replication is again controlled by topoisomerases. 

\begin{figure}[t]
	\centering
	\includegraphics[width=0.45\textwidth]{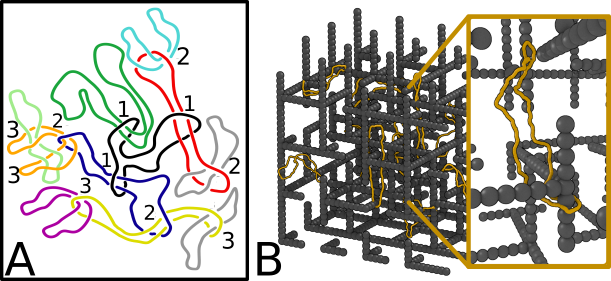}
	\caption{Examples of entanglement, or topological interactions between ring polymers (A) or ring polymers and environment (B), which cannot be captured by a simple topological invariant, and provide the focus of the current review. Linear (or topologically trivial) polymers, do not experience the same constraints in their motion as ring polymers do in these situations. In (A) we report a possible configuration where multiple threadings (inter-penetration) strongly affect the diffusive motion of the central black ring. In (B) we report a snapshot from Molecular Dynamics simulations of charged ring polymers (gold) dragged upwards by an external electric field and ``impaled'' by the dangling ends of a gel (grey), adapted from Ref.~\cite{Michieletto2014ringsoftmatter}.}
	\label{fig:topint}
\end{figure}

There are however situations in which the notion of entanglement is not directly related to a well defined topological state, such as a knot or a link, but is still extremely relevant in affecting the physical properties of the system. Architecture-specific entanglements, dubbed ``threadings'', have been recently shown to proliferate in dense solutions of unlinked and unknotted ring polymers and in the case of dilute rings in disordered environments (see Figure~\ref{fig:topint}). Even more importantly, these threadings have been shown to entail notable consequences on the rings' dynamics. For these reasons, the definition, identification and quantification of these dynamical effects is the focus of the present topical review.

The first example is one of mutual threading in a concentrated solution, or melt, of ring polymers (see Fig.~\ref{fig:topint}A). The threadings (see Section 2 for a more formal definition) in the configuration shown in the Figure are hierarchical, and are numbered from the innermost to the outermost. For instance, the black ring in the centre is threaded by three other rings (threadings numbered as ``1'' in the Figure). The concept of threading is directional, for instance the black ring is (passively) threaded by three rings, whereas the purple ring (actively) threads the yellow ring.  Since rings cannot cross one another (or self-cross), there is no knot or link at any time in the system, so the global topology will be described as trivial by using standard topological invariants (such as, e.g., knot or link polynomials).
However, it is intuitive that the threadings will affect the physical properties of the system. For instance, the mobility of the black ring is severely affected: before it can move away diffusively, all its three passive threadings need to be resolved (and resolving complex threadings such as that with the blue ring will likely take a long time). In other words, the threadings act as effective topological constraints that strongly affect the dynamics of the system. 

Imagine now cutting each of the strings in Figure~\ref{fig:topint}A, so as to create a melt of linear (rather than ring) polymers. The resolution of entanglements can now be faster, as linear polymers can slide past each other and will not (or barely) be hindered by the threadings. Indeed, the dynamics of linear polymers in a melt is well understood and can be described in terms of the tube model and reptation theory. In stark contrast, the dynamics of ring polymers, which as expected from our simple argument can be shown to be very different~\cite{Halverson2011c,Halverson2011d,Kapnistos2008}, is still poorly understood theoretically: one reason why this is so is because entanglements such as the threadings in Figure~\ref{fig:topint}A are so difficult to characterise quantitatively. 
Here we report on a recent method to unambiguously identify and quantify inter-ring threadings in a system of rings embedded within a regular 3D cubic mesh (a simple model for a gel). The presence of a background regular structure is useful in this context as it allows a mathematically rigorous definition of threadings (see Section~\ref{s:detectingthreadings}).

A different example of common physical entanglement, this time between a polymer ring and an obstacle in the environment, is shown in Figure~\ref{fig:topint}B. Here we consider the case where a system of rings is being dragged through an irregular mesh of rod-like obstacles (a more realistic model of a physical gel). A possible experiment recreating the situation shown in Figure~\ref{fig:topint}B is one in which a diluted solution of ring polymers, such as negatively charged DNA plasmids, are forced to move through an agarose gel (which normally contains dangling ends) by an external electric field.
In this case, we expect the defects in the gel (the dangling ends) to topologically interact with the moving rings, for instance creating ``impalements'', between a ring  and a dangling end. Like the configuration in Figure~\ref{fig:topint}A, this entanglement will also affect the ring dynamics, while the topological invariant will always detect no links between the rings and the mesh they travel through.

A method of detecting impalaments and quantifying their statistics has been recently introduced in~\cite{Michieletto2014ringsoftmatter} and used to understand the irregular migration speed detected in experiments comparing linearised and circular plasmids~\cite{Mickel1977,Levene1987}. This method as well as its implication on the rings dynamics will be discussed below.
Quite remarkably, quantifying the level of entanglement between knotted rings and a gel-like environment as that of Figure~\ref{fig:topint}B provides an explanation for the experimental observation that the mobility of DNA knots knotted in 2D gel electrophoresis under strong fields depends nonmonotonically on the knot complexity (a feature that is absent if the model gel displays no dangling ends so that no impalaments can occur, see Section ``\emph{2D Gel Electrophoresis of Knotted DNA}'').

Before proceeding, we stress that, although the entanglement we are discussing here is not strictly topological, the methodology introduced to detect and quantify this property is strongly inspired by topology and in particular it makes substantial use of the notion of linked state and linking number between pairs of rings.
 

\section{Developing Algorithms to Detect Threadings}

In this Section we discuss some recent algorithms which have been developed to identify physical entanglements such as threadings. First, we will consider threadings corresponding to mutual entanglement, between ring polymers (as in Fig.~\ref{fig:topint}A). Then, we will consider impalaments and entanglements between a ring polymer and a gel-like environment (modelled as a cubic mesh, as in Fig.~\ref{fig:topint}B).

\subsection{Detecting Threadings between Ring Polymers}
\label{s:detectingthreadings}

To define and unambiguously quantify  the degree of threading or interpenetration of a concentrated solution of rings is challenging and it has been independently attempted by several groups in recent years~\cite{Smrek2016,Lang2013,Lee2015,Tsalikis2014,Tsalikis2016, Halverson2011c,Halverson2011d}. One of the problems is that, as shown in Figure~\ref{fig:topint}A, threading may involve several rings at once -- as we shall see, the threading network may even percolate through the whole system.
Another difficulty is that, as mentioned, these physical entanglements are not permanent and occur between unlinked rings, so that topological invariants cannot be used in a straightforward manner, as in the case of knotted or catenated polymers~\cite{Stas,Tubiana2011}. 

\begin{figure*}[t]
	\centering
	\includegraphics[width=0.6\textwidth]{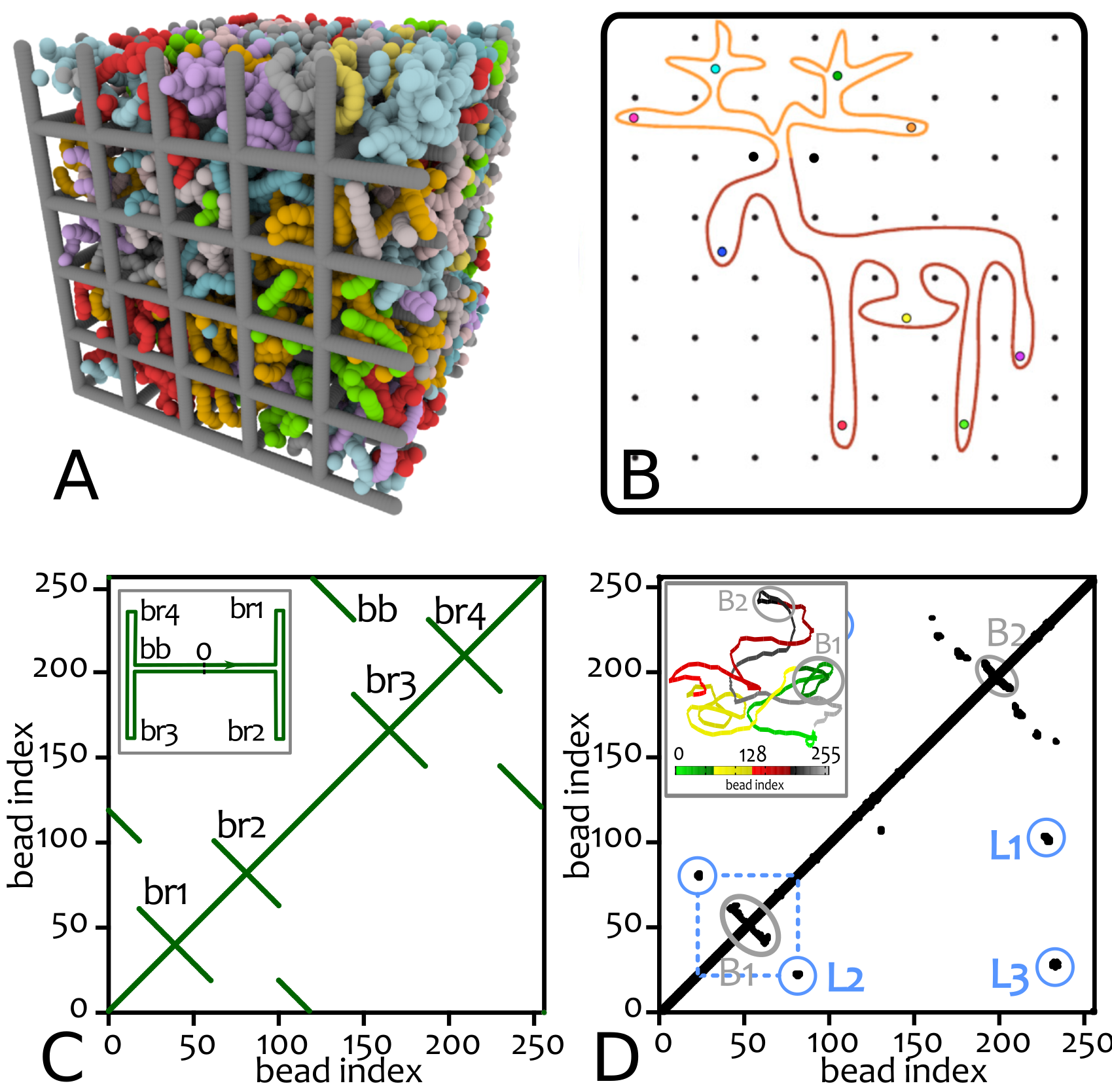}
	\caption{Rings as ``loose'' lattice animals.  (A) Snapshot from Molecular Dynamics simulation representing the dense solution of rings contained within a background mesh of static obstacles reminiscent of an idealised agarose gel. (B) A possible configuration of an isolated ring, assuming a so-called ``lattice animal''~\cite{Lubensky1979} (an elk, in this case) state (from Ref.~\cite{Michieletto2016pnas,Kapnistos2008}). Each black dot represents a un-crossable obstacle, i.e. either a strand of the gel or a neighbouring ring that is not threading through the contour of the animal. Coloured circles instead represent ingoing and outgoing threading sites. (C) Contact map for an ideal lattice animal (sketched in the inset). Notice the sole presence of lines capturing the tightly double-folded nature of the animal. (D) Contact map for a ring polymer in a dense solution (3D configuration reported in the inset and colour coded as per bead index, from Ref.~\cite{Michieletto2016pnas,Michieletto2016treelike}). Notice that the real diagram shows the presence of spots rather than lines, indicating the existence of long-range loops (e.g. L1, spanning about  half contour length), which are good candidate structures to accommodate threadings. }
	\label{fig:ringsingel}
\end{figure*}

Due to these issues, we recently proposed a strategy to identify threadings in dense solutions of polymers embedded in a static 3D cubic lattice~\cite{Michieletto2014}. The gel structure has a fixed lattice spacing $l$, which in Ref.~\cite{Michieletto2014} was taken equal to the Kuhn length of the polymer, which represents the natural length-scale at which the polymers become flexible. 

\begin{figure}[t]
 	\centering
 	\includegraphics[width=0.45\textwidth]{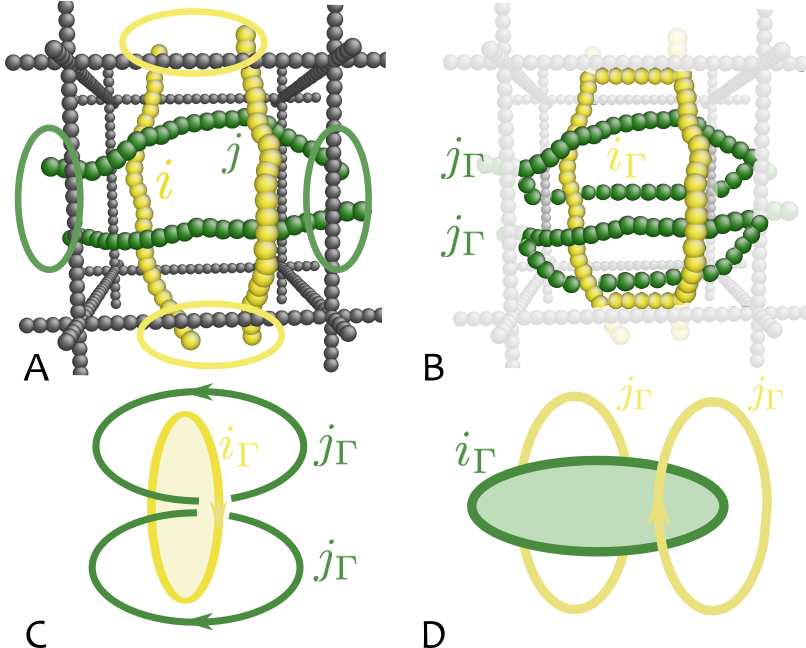}
 	\caption{A strategy to identify threadings.  (A) Consider a pair of rings sharing a unit cell of the gel and identify the points where contours cross the edges of the cell.  (B) Close the contours asymmetrically: the first (yellow) by following the shortest path between exit points. The other (green) strands by extending their paths away from the cell and by joining each strand separately. (C) The resulting situation allows one to identify threadings: the two new rings originating from the second (green) ring are linked to the first new contour if, and only if, the first ring is threaded by the second. (D) In all other cases, the new contours are not linked with one another (from Ref.~\cite{Michieletto2014}).}
 	\label{fig:closureprocedure}
 \end{figure}

The gel structure is fixed in space and time and it only acts an additional set of background obstacles for the rings in solutions -- from which they are all unlinked (see snapshot in Fig.~\ref{fig:ringsingel}A).  This structure is useful as it provides a natural local volume -- a single unit cell. Because rings are unlinked from the cubic lattice, each unit cell $\Gamma$ must display an even number of intersections between its surface and any of the polymers (see Fig.~\ref{fig:closureprocedure}(A)).

Threading of yellow polymer $i$ by green polymer $j$ within cell $\Gamma$ can then be defined as follows. First, a contraction of ring $i$ is formed by sequentially connecting the points where it passes through any face of cell $\Gamma$ by the shortest path . This creates a closed loop $i_\Gamma$ contained entirely within the cell and its bounding faces. Next we consider each of the strands belonging to ring $j$ and we join the ends of each strand outside the cell to form a closed loop labelled $j_\Gamma$ (see Fig.~\ref{fig:closureprocedure}(B)). 

Finally, we compute the pairwise linking number between each $j_\Gamma$ and $i_\Gamma$. The sum over all strands $j_\Gamma$ will return the ``threading number'' $Th_\Gamma(i,j;t)$ as
\begin{equation}
	Th_\Gamma(i, j; t) = \dfrac{1}{2}\sum_{j_\Gamma} \left| Lk(i_\Gamma, j_\Gamma; t) \right|.
	\label{eq:threadingnumber}
\end{equation}
This quantity will be equal to $1$ if, and only if, ring $i$ is threaded once by ring $j$ in cell $\Gamma$ (see Fig.~\ref{fig:closureprocedure}(C)). Otherwise, reversing the role of chains $i$ and $j$, and following the same procedure (i.e., first contracting ring $j$ and then joining each of the two strands of polymer $i$ outside $\Gamma$) will lead to no linking, i.e. $Th_\Gamma(j, i; t)=0$ (see Fig.~\ref{fig:closureprocedure}(D)). Similarly, it can be checked that this construction leads to non-linking contours also when chains sharing the same call are not threading at all. Finally, it is worth notice that the matrix elements, $Th_\Gamma(i,j;t)$ and $Th_\Gamma(j,i;t)$, need not to be equal; on the contrary, $Th_\Gamma(i,j;t)=1$ if $i$ is threaded by $j$ whereas $Th_\Gamma(j,i;t)$ may be zero, meaning that $j$ is not threaded by $i$, thereby formally identifying $i$ as ``active'' trheading and $j$ as ``passive'' threaded ring.

This strategy works also for more complicated contours and it results in an unambiguous identification of \emph{local} threading between chains. Finally, the total number of threadings between any two rings can be obtained by summing over all cells $\Gamma$, i.e. $Th(i,j;t)=\sum_\Gamma Th_\Gamma(i,j;t)$.

\subsection{Quantifying Entanglement of Rings with a Disordered Environment}
\label{s_gelentanglement}

We now consider the case in which ring polymers are in a disordered environment such as that of a real agarose gel~\cite{Cole2002} or network of nanowires~\cite{Rahong2014}. We wish to characterise the topological interactions between rings and environment, which can be crucial to determine their mobility. As an example of an observable consequence of these interactions, we will discuss in Section~\ref{knots} the electrophoretic mobility of DNA knots within an agarose gel~\cite{Stas,Katritch1996a}. Experiments in the past have also shown that ring polymers molecules can be ``trapped'' by the dangling ends populating an agarose gel~\cite{Cole2003a,Guenet2006}. These interactions correspond to the impalements shown in Figure~\ref{fig:topint}A, and in this Section we show how to quantify the statistics (i.e., the strength and frequency) of impalements as a function of knot complexity.

\begin{figure*}[t]
  \centering
  \includegraphics[width=0.9\textwidth]{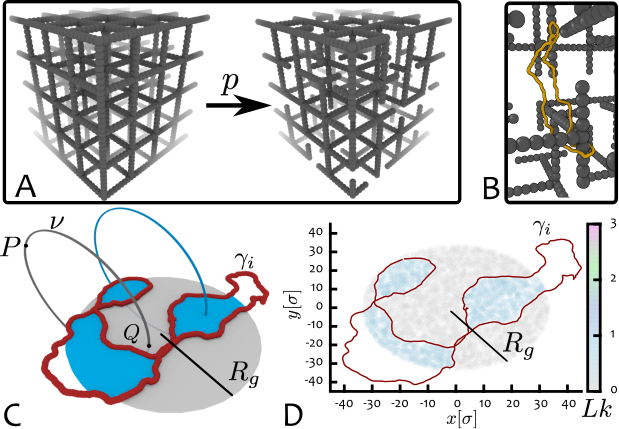}
  \caption{Quantifying impalements of rings with disordered gels. (A) An agarose gel is modelled as an imperfect cubic lattice where half-edges are removed with probability $p$. This disordered gel is constructed in such a way it possesses dangling ends~\cite{Cole2002,Cole2003a} (from Ref.~\cite{Michieletto2016phdthesis}). (B) Dangling ends can trap migrating rings (by forming impalements). (C) Equilibrated ring configurations $\gamma_i$ (for ring polymer $i$) are projected onto a plane and the area corresponding to a disk of radius $R_g$ sampled using virtual closed paths $\nu$. These start from different points $P$ far away from the projections and pierce the disk once at different locations $Q$. (D) The average piercing number $\langle \pi \rangle$ is then computed averaging the value of the linking number, $Lk(x,y; \gamma_i,\nu)$ (shown as density map in the figure), over different projections of $\gamma_i$ and sampling different virtual paths $\nu$ over the area of the disk (from Ref.~\cite{Michieletto2015pnas}). The average piercing number gives an estimation of how likely an entanglement between the ring and a dangling end is.}  
  \label{fig:piercing}
\end{figure*}
  
A possible strategy to quantify the likeliness of impalements is described in Figure~\ref{fig:piercing}~\cite{Michieletto2015pnas}: the quantity measuring this likeliness is the ``piercing number''. The idea to compute the piercing number is similar in spirit to the one used in knot theory to find the average crossing number (ACN) of a ring polymer $\mathcal{K}$, which can be defined as~\cite{Stas}
\begin{equation}
	ACN(\mathcal{K})=\dfrac{1}{4\pi} \int_{ \gamma_\mathcal{K}} \int_{ \gamma_\mathcal{K}} \dfrac{ \left| \left( \bm{r}(s) - \bm{r}(t) \right) \cdot \left( d\bm{r}(s) \times d\bm{r}(t) \right) \right|}{\left| \bm{r}(s)-\bm{r}(t)\right|^3},
	\label{eq:acn}
\end{equation}
where $\bm{r}$ is the position vector describing the ring polymer, with contour $\gamma_\mathcal{K}$.

\begin{figure*}[t]
  	\centering
  	\includegraphics[width=0.95\textwidth]{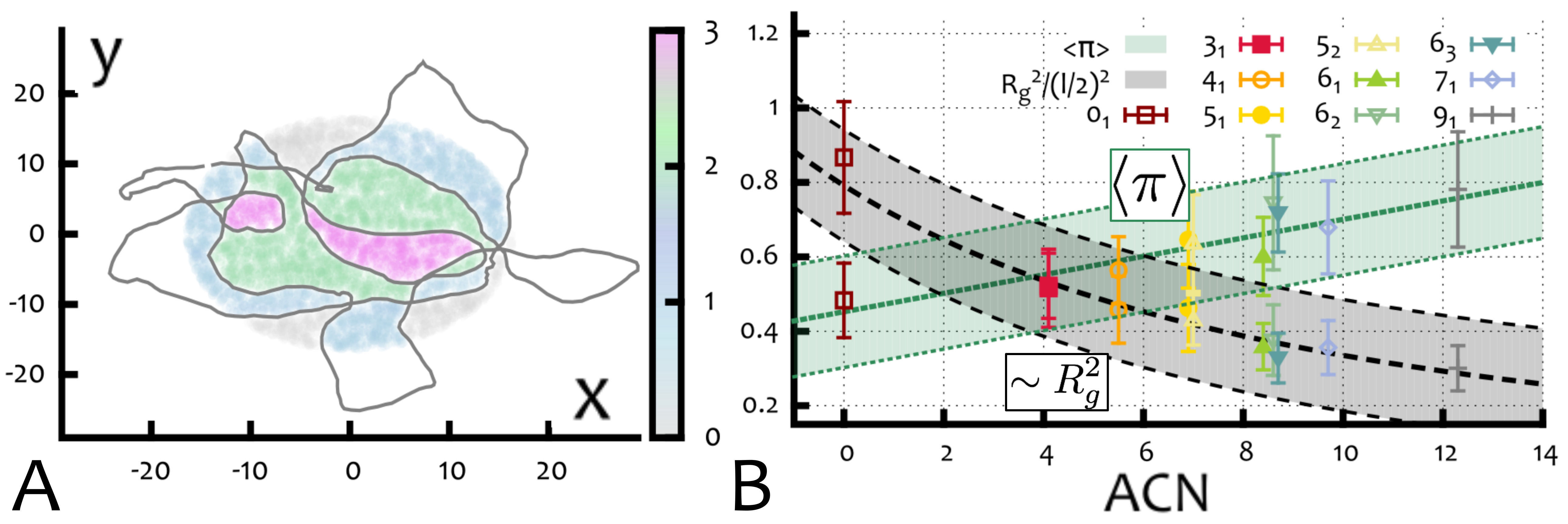}
  	\caption{\textbf{(A)} Entanglement field $\pi(x,y)$ over a disk $\mathcal{D}$ for a $9_1$ knot whose projection is traced as a grey curve. \textbf{(B)} Entanglement number $\langle \pi \rangle$ and radius of gyration (squared) $R^2_g/(l/2)^2$ as a function of the average crossing number $ACN$. Here $l$ is the lattice spacing of the model gel. Data points correspond to the value taken by various knot types ranging from $0_1$ to $9_1$.}
  	\label{fig:ent_numb}
  \end{figure*}

Specifically, to compute the piercing number we use the following algorithm. First, we take equilibrated configurations of ring polymers, project them onto a plane and define a disk region $\mathcal{D}$ of radius $R_g$ lying on the same plane of the projected contour, and centred in its centre of mass (as in Fig.~\ref{fig:piercing}(C)). Second, we generate many virtual closed paths $\nu$ starting at points P away from the polymer and passing (piercing) through the disk $\mathcal{D}$ exactly once. Third, we compute the linking number between each of the closed paths and the projected ring configuration and assign the value of $Lk(x,y; \gamma_i,\nu)$ to the intersection $(x,y)$ between $\nu$ and $\mathcal{D}$. By averaging over many possible $\nu$, one can generate the piercing number, a scalar field  $\pi(x,y)$ over the disk $\mathcal{D}$ (see Fig.~\ref{fig:piercing}(D) and \ref{fig:ent_numb}(A)). The average of the field over the area of the disk and over different projection planes and equilibrium configurations can be defined as the ``average piercing number'' $\langle \pi \rangle$, which gives a measure of the extent of possible entanglement between the ring $\mathcal{K}$ and the disordered environment.  

It is clear that this procedure is an arbitrary construction and it is all but the only one possible. On the other hand it has several advantages. First, the creation of virtual paths piercing through the projected contour of a polymer conformation mimics the piercing of dangling ends through the contour of a molecule when moving through a gel. On can thus imagine that for a given ring projection, each region of the corresponding disk $\mathcal{D}$ is assigned with a field $\pi(x,y)$ that quantifies the complexity of a possible impalement through that region. Second, the size of the disk $\mathcal{D}$ is equal to $R_g$, therefore effectively taking into account the fact that more tightly folded rings (e.g., some complex knots) offer less ``threadable'' area. The average piercing number is the first observable, to our knowledge, that can measure the complexity of an entanglement of a ring with a disordered environment. Its similarity to the more standard average crossing number -- used to measure the ``self-entanglement'' of a molecule -- is that it too can be interpreted as the result of an averaging over a large number of projections of the ring $\mathcal{K}$. 

For knotted rings, we find that the average piercing number $\langle \pi \rangle$ increases linearly with the average crossing number of a knot (Fig.~\ref{fig:ent_numb}(B)). Additionally, by comparing the behaviour of $\langle \pi \rangle$ and $R_g$ (Fig.~\ref{fig:ent_numb}(B)), one can visualise the competition between size of a knot (and therefore its ``threadable'' area) and its complexity (and therefore the degree of complexity of a potential entanglement). In Section~\ref{knots} we will show that this competition give rise to non-trivial macroscopic behaviour that can be seen in 3D gel electrophoresis experiments. 
  
\section{Results}  
In this Section, we review the main findings and observations which were obtained by using the methods described above to describe entanglement and threading in systems of ring polymers. 

\subsection{The Statistics of Threading in Ring Polymers}
\label{threadingstat}

As the concentration of ring polymers in a melt increases, the gyration radius of individual rings, $R_g$, scales with polymerisation index $M$ as $R_g \sim M^\nu$ with the \red{exponent in the range $ 1/3 \lesssim \nu = 0.4$ for large $M$}~\cite{Grosberg1993,Halverson2011c,Cates1986,Nechaev1987,Sakaue2011,Rosa2013, Nahali2016,Muller2000}. A long-standing question in polymer physics is whether this compaction ($\nu=0.588$ for swollen, or isolated, rings) is accompanied by a decrease (increase), of mutual topological entanglement (number of threadings) between the rings.  

\begin{figure}[t]
	\centering
	\includegraphics[width=0.40 \textwidth]{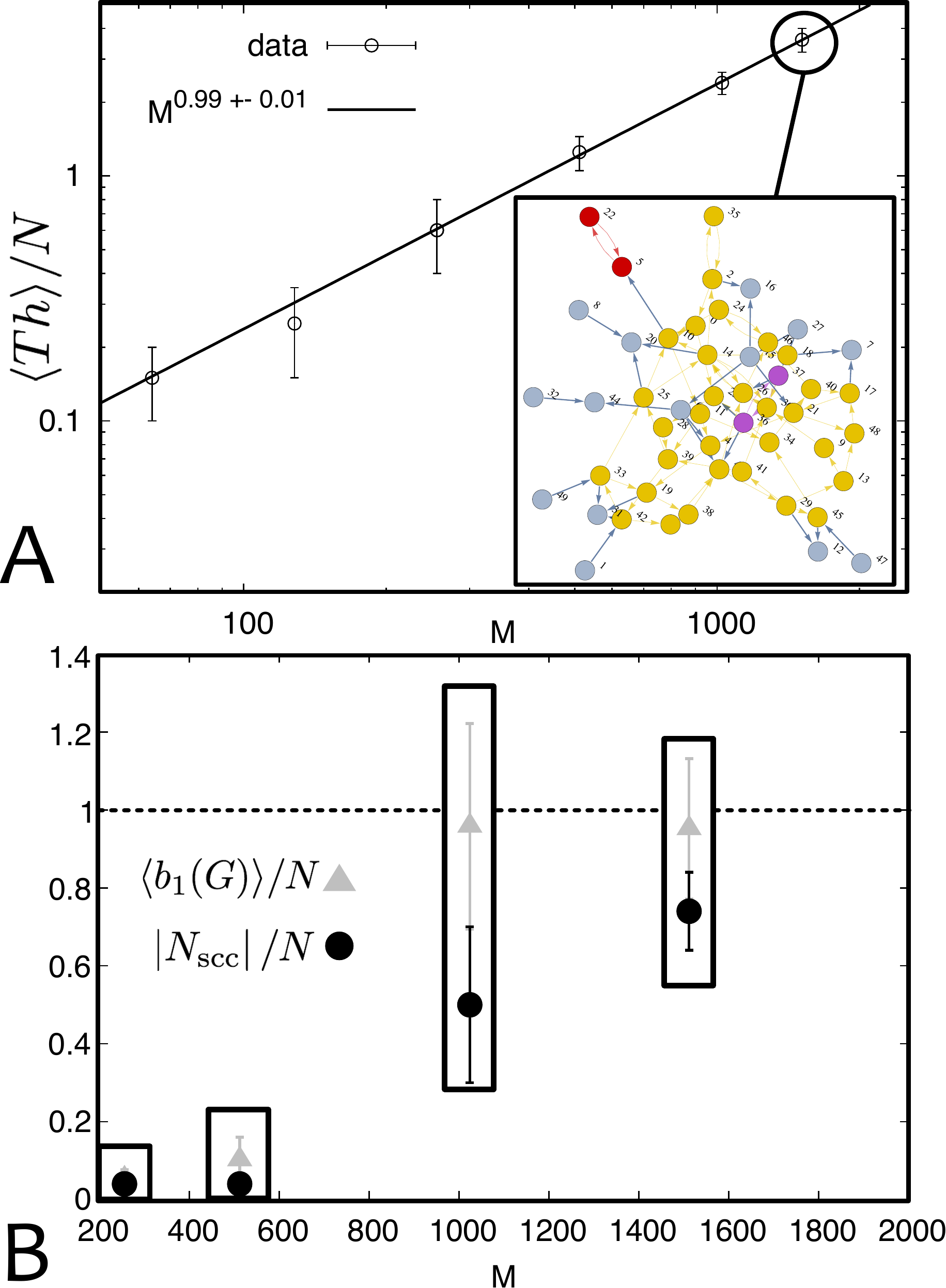}
	\caption{\textbf{(A)} Threadings can be directly detected through the algorithm described in the text. The number of threadings in the system is found to grow linearly with the size of the rings. (Inset) The proliferation of threadings can be seen as forming a network of constraints, here captured by a graph $G$ with directed edges pointing to actively threaded rings. \textbf{(B)} The network can be quantified by measuring the Betti number $\langle b_1(G) \rangle$ and the size of the largest strongly connected component $\left|N_{\rm scc} \right|$. These are observed to sharply increase around $M \simeq 1000$ and the fraction of rings $\left|N_{\rm scc} \right|/N \rightarrow 1$ at large $M$, indicating that the network of threadings percolates through the whole system.}
	\label{fig:n_th}
\end{figure}


Thanks to the algorithm described in Section ``\emph{Detecting Threadings between Ring Polymers}'', we can address this question by identifying threadings between any two pair of polymers in a system of rings embedded in 3D lattices. 
We expect that the trends observed for the number of threadings will be qualitatively similar to the case of a pure melt of rings, and as we shall see, computer simulations indeed support this expectation (see also Ref.~\cite{Smrek2016}). Thus, we performed Molecular Dynamics (MD) simulations (see Suppl. Material for computational details) of solutions of rings at fixed monomer density $\rho\sigma^3=0.1$ and with chain length in the range $M=256-1024$. The average number of threadings in the system  $\langle Th \rangle$ can be obtained through the time-dependent adjacency matrix $Th(i,j;t)$ (see Eq.~\ref{eq:threadingnumber}), and numerical results show~\cite{Michieletto2014}
\begin{equation}
	\langle Th \rangle/N \sim M^\alpha
	\label{eq:thvsm}
\end{equation} 
with $\alpha=1$ within errors (Fig.~\ref{fig:n_th}(A)). Wheras many works in the literature conjecture the presence of threadings~\cite{Muller1996a,Muller2000,Halverson2011c,Lee2015,Lang2013,Rosa2011,Rosa2013,Tsalikis2014,Tsalikis2016}, they lack a dedicated algorithm to give quantitative estimations for the number of \emph{ring-ring inter-penetrations}. To our knowledge, the only other quantification for this observable in melt of rings has been obtained in Ref.~\cite{Smrek2016} and it fully agrees with our original finding, i.e. that $\langle Th \rangle/N$ scales linearly with the size of the rings $M$. 

Eq.~\eqref{eq:thvsm} states that the number of constraints proliferates with the rings polymerisation index $M$, thereby suggesting that, although rings become more compact (\red{$1/3 \gtrsim \nu = 0.4$} in the melt~\cite{Grosberg1993,Halverson2011c,Grosberg2013,Rosa2011,Rosa2013,Nahali2016}) they also inter-penetrate much more. This picture can be recapitulated by quantifying the adjacency matrix $Th(i,j;t)$. In the inset of Fig.~\ref{fig:n_th}(A), we show a typical snapshot of the network of threadings, which is represented by a directed graph $G$, where nodes represent rings and arrows pointing from $i$ to $j$ indicate that ring $i$ is threading through ring $j$. As one can readily notice mutual threadings ($i \leftrightarrow j$) can also appear. 

A graph representation of the threading network provides a natural mean to visualise the structure of these constraints in systems of ring polymers. To analyse the topology of this network, we analysed this graph to extract: (i) the size of the largest (strongly) connected component $\left| N_{\rm scc} \right|$ (i.e., the size of the subset of nodes that can be reached by following the edges and starting from any of the nodes belonging to that subset~\cite{newman2010}), and (ii) the first Betti number $b_1(G)$, which measures the size of 1D ``holes'' in the graph~\cite{newman2010}. As one can see from Fig.\ref{fig:n_th}(B), both quantities are seen to sharply increase around $M \simeq 1000$ and eventually attaining the value of $N$ (maximum number of rings in the system) at large polymerisation indexes $M$.  

The numerical results just described suggest the following physical picture. If we consider a system of rings at fixed volume fraction $\Phi$ (within a gel), increasing the contour length $M$ of the rings leads to a transition, or crossover, between a regime where threadings are rare and the network of inter-ring penetration is sparse and disconnected, to another regime where entanglements are ubiquitous and a fully connected network of threadings percolates through the whole system. It is natural to expect that these percolating network of topological interactions will likely leave a distinctive signature in the dynamics of the rings. For instance, it is intuitively clear that extracting a single ring (or rubber band) from a melt in the percolating regime will be extremely hard -- as all the other rings will be dragged with it due to the entanglement~\cite{physworld2014}. Therefore, in the next Section we quantitatively discuss the dynamical consequences of threadings. 

\subsection{Polymer Melts and Topological Glasses}  
\label{topologicalglass}

In this Section, our goal is to characterise the way in which threadings affect the dynamics of rings. This is an important problem that has been the focus of several groups in recent years~\cite{Lee2015,Tsalikis2016,Vlassopoulos2016,Smrek2016,Milner2010,Ge2016,Smrek2015} and that has profound consequences for the macroscopic and rheological behaviour of polymer melts with closed topology~\cite{Kapnistos2008,McLeish2002,McLeish2008,Vlassopoulos2016}. We argue~\cite{Michieletto2014,Lo2013} that threadings lead to a dynamical slowdown, which eventually may result in a kinetical arrest, of the polymer dynamics (the dynamical slowdown due to threadings has also been subsequently argued in Refs.~\cite{Lee2015,Tsalikis2016,Smrek2016}). This slowdown is to some extent reminiscent of the one observed at or near the glass transition which occurs in a suspension of hard colloidal spheres (or polymers) in water as their packing fraction increases past a threshold~\cite{Jones} -- although in this case the mechanism underlying the glassy dynamics is purely topological (the proliferation of inter-ring threadings). 

To characterise the ring dynamics, we employ the strategy of perturbing the system to study its dynamical response. In analogy with classical ``thought experiments'' for systems of linear and ring polymers~\cite{Doi1988,Gennes1979,Rubinstein1986} and with more recent work on the glass transitions~\cite{Cammarota2012}, we consider systems where a random fraction of constituents is frozen in space and time. Therefore, a fraction of our rings is frozen, whereas the rest can move. This set-up can be experimentally realised for instance, by mixing two types of ring polymers having glass temperatures $T_{g,1}$ and $T_{g,2}$ respectively, and by setting the solution temperature $T$ such that $T_{g,1} < T < T_{g,2}$~\cite{Lodge2000}. Under this condition, the first type of polymers will remain mobile, whereas the second one will be virtually frozen. Another possibility to recreate our setup in the lab is to use optical tweezers~\cite{Gokhale2014} to ``pin down'' some of the polymers (this is similar to the strategy used in colloidal systems). Finally, in the future it might be possible to design ``topological probes'' which can be employed in solution of rings to link and immobilise a subset of the polymers.  
Note that, in all the simulations reviewed in this Section, there is no longer a background gel.
   
To begin with, we report results from molecular dynamics simulations~\cite{Michieletto2016pnas} for the case in which all the ring polymers are pinned (i.e. permanently immobile), apart from one, which is left free to move. In addition, to identify more clearly the role of topology, we compare the results with the case of a single linear chain diffusing through a frozen background of pinned linear polymers. For reference, we also consider the cases of a ``standard'' molecular dynamics simulation, where all polymers are mobile -- again, considering the cases of rings and linear chains separately. 

In each of these cases we analyse the behaviour of the mean square displacement of the centre of mass of the mobile polymers as a function of the lag-time $t$, 
\begin{equation}
	g_3(t) = \langle \left| \bm{r}_{CM}(t_0) - \bm{r}_{CM}(t_0+t) \right|^2 \rangle\, ,
\end{equation}
where $\bm{r}_{CM}(t)$ is the position of the centre of mass t time $t$ and $\langle \, \dots \rangle$ denotes time and ensemble average. 

\begin{figure}[t]
	\centering
	\includegraphics[width=0.45 \textwidth]{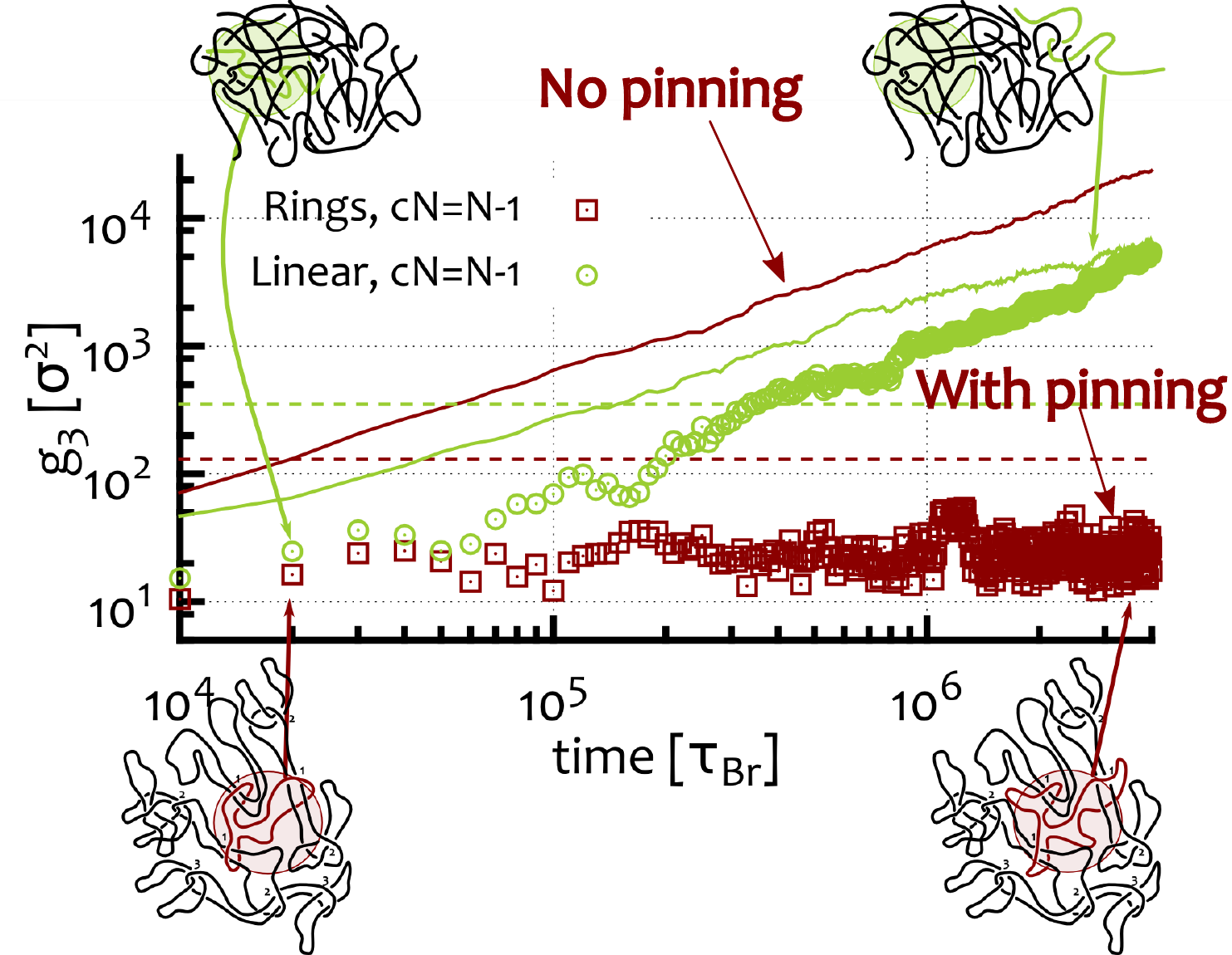}
	\caption{Comparison between the mean square displacements of the centre of mass of the mobile polymers in four cases. Open green circles: a mobile linear chain in a background of pinned (frozen) linear chains. Open red squares: a mobile ring polymer in a background of pinned (frozen) ring polymers. Green line: a melt of linear polymers. Red line: a melt of ring polymers. All systems have the same monomer density $\rho=0.1 \sigma^{-3}$, and include $N=50$ polymers, each of length $M=256$ beads.}
	\label{fig:g3_linvsrings}
\end{figure}

The curves obtained for the four cases (linear and rings, where all but one polymers are frozen, or where they are all mobile) can be compared and contrasted in  Figure~\ref{fig:g3_linvsrings}. The case of a free linear polymer probe within a system of pinned linear chains (open green circles) shows that, at long times, the diffusion coefficient is similar to that observed in the ``standard'' linear case (where all polymers can move, solid green line in Fig.~\ref{fig:g3_linvsrings}). Therefore, the dynamics is not affected by the pinned background in the long time limit (the initial slowdown in the curve is caused by the infinitely slow relaxation of the nearest neighbours and therefore of the initial ``tube''~\cite{Doi1988} due to the pinning). In marked contrast with this behaviour, the diffusion of a single probe ring in a background of frozen rings is heavily suppressed ($g_3 \sim t^0$, open red squares in Fig.~\ref{fig:g3_linvsrings}), so that pinning now has an enormous effect on the dynamics.

These findings clearly indicate that the closed topology of the rings and their architecture-specific topological constraints (i.e., inter-ring threadings) dramatically affect their dynamics in solutions perturbed by random pinning. We now ask whether similar effects can be seen in the limit of no pinned ring, i.e. in unperturbed solutions. To address this question we report results from Ref.~\cite{Michieletto2015pnas}, which studied the response of systems of rings to random pinning perturbations which explicitly freeze only a fraction $c$ of rings. Eventually, we will be therefore interested in the limit $c\rightarrow 0$. As mentioned before, while this strategy has been inspired by recent works on the study of the glass transition in colloidal systems subject to random pinning~\cite{Cammarota2012,Karmakar2013,Nagamanasa2011,Nagamanasa2014,Gokhale2014,Weeks2015,Kob2014a,Chakrabarty2014}, it has never been employed, to our knowledge, to directly study the dynamical behaviour of systems of polymers.  

\begin{figure*}[t]
	\centering
	\includegraphics[width=0.85 \textwidth]{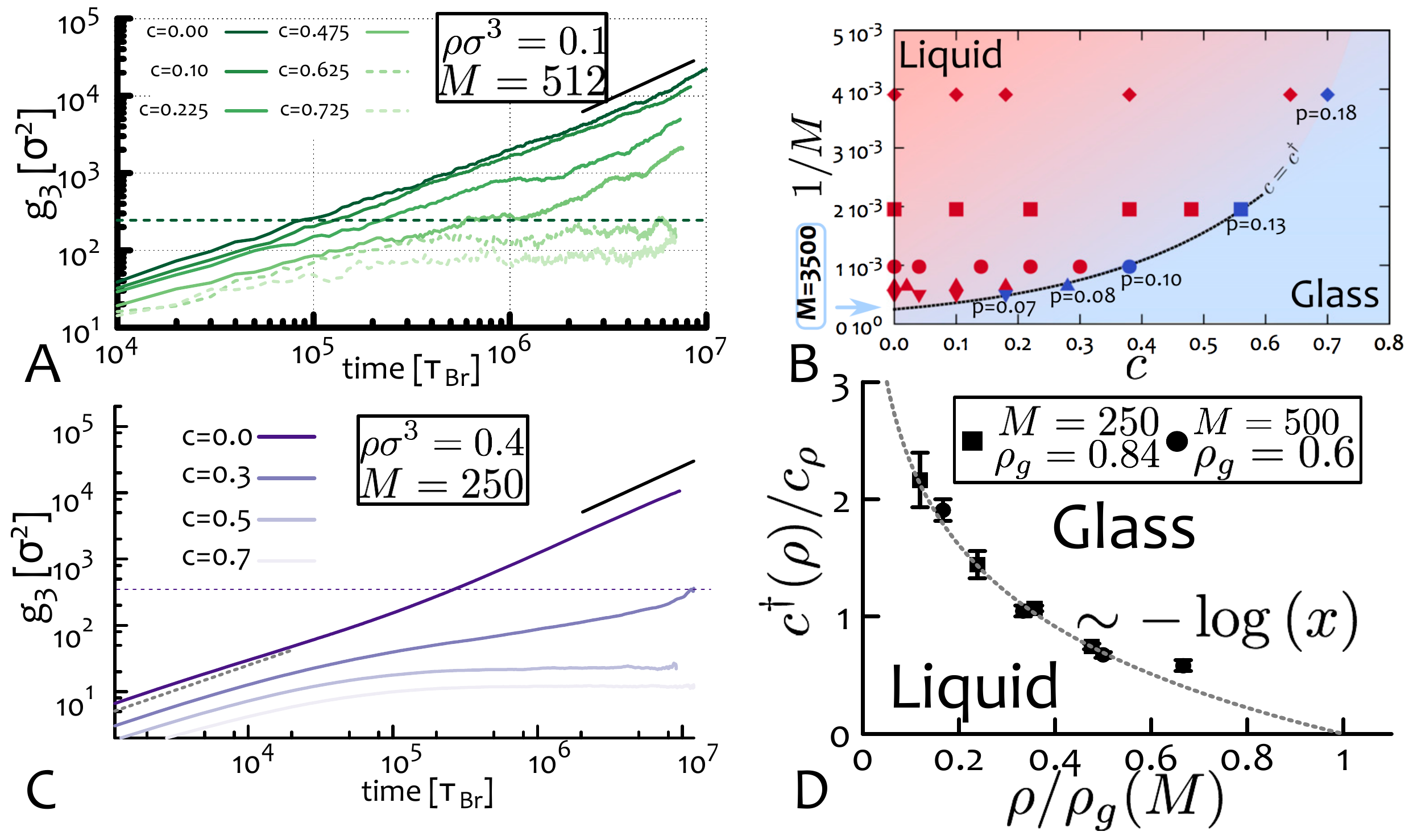}
	\caption{\textbf{(A)} Transition from liquid to glassy behaviour as a function of increasing pinned fraction $c$ and for a system of rings with $M=512$ and $\rho=0.1 \sigma^{-3}$. for which we find $c^\dagger \simeq 0.6$. \textbf{(B)} Phase diagram of the system in the parameter space $(c,1/M)$. The ``critical'' line $M^\dagger(M)=M_g\exp{(-c/c_M)}$ separates regions with finite (liquid) and vanishing (glassy) diffusion coefficients. The values of $p$ along the transition line represent the probability of finding an uncaged ring in any one system replica as computed through a binomial distribution at 95\% confidence interval, i.e. $(1-p)^{(1-c\dagger)N}=0.05$. \textbf{(C)} Transition from liquid to glassy behaviour as a function of increasing pinned fraction $c$ for system of rings with $M=250$ and $\rho=0.4 \sigma^{-3}$. In this case we find $c^\dagger(\rho) \simeq 0.32$. (D) Phase diagram of the system in the parameter space $(\rho,c)$. Rescaling the values of $c^\dagger$ by a  constant $c_\rho \simeq 0.43$ and $\rho$ by $\rho_g(M)$ we find a collapse of the data points onto a master curve $c^\dagger/c_\rho = - \log{\left(x = \rho/\rho_g(M)\right)}$ (see Eqs.~\eqref{eq:cdag_M}-\eqref{eq:cdag_rho}).}
	\label{fig:phasediag}
\end{figure*}

In Figure~\ref{fig:phasediag}(A) we show an example where we consider a system of $N=60$ rings $M=512$ beads long and monomer density $\rho=0.1 \sigma^{-3}$. Starting from $c=0$ (unperturbed solution) we increase the value of the pinned fraction until we observe that for $c=c^\dagger(M)$, the curve $g_3(t)$ fails to overcome the average ring diameter ($2 R_g$, dashed line). To more accurately extract $c^\dagger$, it is possible to fit the behaviour of the large time diffusion coefficient (which can in principle depend on polymer length $M$ and system density $\rho$)
\begin{equation}
	D(M,\rho; c)=\lim_{t \rightarrow \infty} \dfrac{g_3(t)}{6t}
\end{equation}  
with exponential functions $d(c)=\exp(-k c)$, and by finding the intersection of $d(c)$ with a threshold, as done in~\cite{Michieletto2016pnas} (the threshold there was set to $0.02$ -- the value of $c^\dagger(M)$ in Fig.~\ref{fig:phasediag} is estimated in this way). 

By repeating these simulations for solutions of rings with length $M$ and density $\rho$~\cite{Michieletto2017}, we can map out the phase diagrams in Figure~\ref{fig:phasediag}B and Figure~\ref{fig:phasediag}D, where liquid and glass regimes are defined in terms of their long-time diffusion coefficients. Simulations show that the following empirical relations hold~\cite{Michieletto2016pnas,Michieletto2017}
\begin{align}
	&c^\dagger(M,\rho) = -c_M \log{\left( \dfrac{M}{M_g}\right)} \label{eq:cdag_M}\\
	&c^\dagger(M,\rho) = -c_\rho \log{\left( \dfrac{\rho}{\rho_g}\right)}. \label{eq:cdag_rho}
\end{align}
In the phase diagrams in Figures~\ref{fig:phasediag}B,D, we also shown the ``critical'' line fitted by the expressions in Eqs.~\eqref{eq:cdag_M}-\eqref{eq:cdag_rho}. We highlight here that Eqs.~\eqref{eq:cdag_M}-\eqref{eq:cdag_rho} describe the same transition line, and note that $c_\rho$ is numerically found to be approximately independent on $M$~\cite{Michieletto2017} (Fig.~\ref{fig:phasediag}(D)). We further assume that $c_M$ is independent on $\rho$ (reasonable in view of the previous results, although numerical data are not currently available to check this), so that all the functional dependencies on $\rho$ and $M$ are contained in $M_g$ and $\rho_g$, respectively. Given these assumptions, we can write
\begin{equation}
 \dfrac{\rho}{\rho_g(M)} = \left( \dfrac{M}{M_g(\rho)}\right)^{\eta} 
\end{equation}  
with $\eta=c_M/c_\rho$~\cite{Michieletto2017}, or 
\begin{equation}
\rho_g(M) = \rho M_g(\rho)^{\eta}  M^{-\eta}. 
\end{equation}  
Finally, we impose that this equation must be valid at any $\rho$, which is equivalent to setting $\rho M_g(\rho)^{\eta}$ to be a constant, or
\begin{align}
&\rho_g(M) \sim M^{-\eta} \notag \\ 
&M_g(\rho) \sim \rho^{-1/\eta}\, . 
\label{eq:scaling}
\end{align}
In the equations above, the exponent $\eta$ takes a numerical value of $\eta=c_M/c_\rho \simeq 0.68 \pm 0.1$ (see Ref.~\cite{Michieletto2017}), where the error originates from statistical uncertainties in the fitting of the $g_3(t)$ curves and, consequently, on the decay of the effective diffusion coefficients $D(M,\rho;c)$.  These scaling relations describe the behaviour of the ``critical'' polymer length $M_g(\rho)$ and system density $\rho_g(M)$ at which spontaneous slowing down due to topological constraints are expected to emerge and may be probed by future experiments testing the dynamics of melts of rings subject to random pinning perturbations. 

It is important to stress that Eqs.~\eqref{eq:scaling} describe the ``mean-field'' response to the pinning perturbation and are based on empirical relations (Eqs.~\eqref{eq:cdag_M}-\eqref{eq:cdag_rho}). In fact, the simulated systems considers only a small part of the true bulk and may thus present some important finite-size effects. 
One problem that is related to the concept of ``percolating network'' of threadings mentioned above is that for a given system configuration, there exist some rings that are more threading than others, i.e. nodes that have more edges. This means that by pinning those rings results in a stronger caging effect. The opposite may also occur, where the pinned rings may display far fewer threadings that the average ring. At present it is difficult to take into account these fluctuations in the theoretical description through which we obtain Eqs.~\eqref{eq:scaling}; nonetheless, an approximate argument for estimating the stochasticity in the behaviour of the rings subject to the pinning perturbation can be described as follows: by applying the same pinning perturbation (tuned by the parameter $c$) to several system replicas, one can estimate the probability $p$ of finding an uncaged ring in any one of the replicas through a binomial distribution at (for instance) 95\% confidence interval as $(1-p)^{(1-c^\dagger)N}=0.05$, where the exponent $(1-c^\dagger)N$ is the number of non-explicitly pinned rings. This equation shows that for large values of $N$ (large system sizes), the probability $p$ of finding a ring that escapes caging due to fluctuations in the number of threadings decreases, whereas for small system sizes the fluctuations become more important (see also Fig.~\ref{fig:phasediag}).  A more thorough analysis of finite-size effects for the case of pinned rings in solution will be addressed in future works. 

A further particularly significant result conveyed by Fig.~\ref{fig:phasediag} is that the transition lines appear to cut through the $1/M$ and $\rho$ axis at finite points, $M_g$ and $\rho_g$. This suggests that a spontaneous (zero pinning fraction) transition from liquid to solid-like behaviour may occur for these semi-flexible rings. For density $\rho=0.1\sigma^{-3}$ this is expected to occur at $M\simeq 3500$ (see Fig.~\ref{fig:phasediag}B) whereas for $\rho=0.4 \sigma^{-3}$, the critical length $M_g$ is expect to decrease to about $500-600$ beads~\cite{Michieletto2017}. Large-scale simulations of systems with these parameters are currently very demanding and will require further extensive work to be investigated in detail.

The glass-like behaviour that is expected to emerge at $c=0$ (or with only very weak perturbations) can be named a ``topological glass'', because it arises far below the standard jamming density and far above the more common glass temperature. Importantly, it \emph{only} relies on the topology of the constituents which impose mutual constraints through an intricate network of threadings, rather than through purely excluded volume interactions as in colloidal glasses.
For this reason, it would be interesting to investigate whether thermodynamic quantities such as pressure, entropy, free energy and indeed the very equation of state of the system display discontinuities or singularities approaching the topological glass point.

To summarise, in this Section we have showed that threadings can be indirectly detected by studying the dynamical response of the system to random pinning perturbations. In particular, we have also shown that the dynamics of ring polymers (but not linear chains) can be slowed down dramatically, or even arrested, by pinning a sufficient number of polymers, and this is due to proliferation of topological entanglements between the rings.

\subsection{Rings in the Melt as Chains with Mobile Slip-links}
\label{sliplink}

One of the outstanding problems within the picture offered up to now is to find a way to reconcile the observation that rings become more and more crumpled as they get longer and denser~\cite{Halverson2011c,Rosa2013,Nahali2016} (i.e., their exponent is \red{$1/3 \gtrsim \nu  = 0.4$}) with the finding that threadings become more and more frequent and long-lived~\cite{Michieletto2014,Michieletto2016pnas,Smrek2016,Lee2015,Tsalikis2016}. At first sight, these two observations appear to contradict one another, since space-filling objects (as compact or crumpled rings with \red{$\nu \lesssim 2/5$} may be) are expected to expel neighbouring chains from their occupied volume. In this Section we discuss a framework, first proposed in~\cite{Michieletto2016treelike} which suggests a possible resolution of this apparent paradox.

\begin{figure}[t]
	\centering
	\includegraphics[width=0.5 \textwidth]{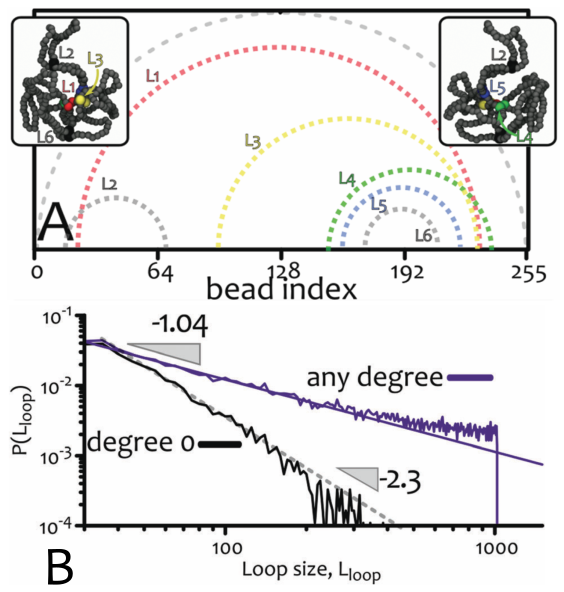}
	\caption{(A) Arc-representation of a ring conformation. An arc is drawn between beads forming a contact, i.e. a loop (L1-L6 in the figure). (B) Probability distribution of loop size length $L_{\rm loop}$. Considering only loops of zero-degree (i.e. not containing other loops), we obtain an exponent $\gamma \simeq 2.3$, close to the one for slip-linked rings~\cite{Michieletto2016treelike} whereas accounting for any loops we recover $\gamma \simeq 1$ as already observed in the literature~\cite{Halverson2011c} and justified by theoretical arguments~\cite{Grosberg2013}. Both curves are normalised by the total number of loops to aid the interpretation and are obtained for rings $M=2048$ beads long. }
	\label{fig:sliplinks}
\end{figure}

In a ring polymer, local proximity between different regions can be monitored by contact maps~\cite{Brackley2016} as those shown in Figure~\ref{fig:ringsingel}. In these maps, a black (or coloured) pixel denote two monomers in the ring which are spatially proximate in 3D (in practice, whose mutual is smaller than a given threshold). 

Within these maps, one can frequently observe looping, rather than double-folded structures, which would be expected if the ring attain branched structures as in Figure~\ref{fig:ringsingel}. This observation suggests that rings in the melt may be thought of as assuming ``loose'' lattice animal conformations. Looping and folding is common in biomolecules~\cite{alberts} and it is typically represented through arc-diagrams for instance in proteins. For ring polymers in the melt, these diagrams (Fig.~\ref{fig:sliplinks}A) show that loops can form pseudo-knots, i.e. arcs cross through each-other~\cite{Michieletto2016treelike}. Within this picture, rings form temporary contacts between beads, which may last as long as they are enforced by surrounding topological constraints (e.g., other nearby rings, or other bits of the same ring, which force the beads into contact locally). Each contact may therefore be thought of as being enforced by a slip-link that is allowed to detach with a rate inversely proportional to the relaxation time of the constraints. At the same time, the total number of slip-links at any given time remains approximately constant over time (as the number of contacts in a map is approximately constant over time in equilibrium).

The statistics of loops and self-contacts have been studied in the literature~\cite{Halverson2011c,Rosa2013}, and recent theories~\cite{Halverson2011c,Grosberg2013,Mirny2011} suggest that the probability of a monomer to contact another bead, $l$-segments apart should scale as 
\begin{equation}
	P_c(l) \sim l^{-\gamma} 
	\label{eq:c_prob}
\end{equation}
with $\gamma \simeq 1$. This exponent has been measured to be in the range $1.02-1.09$ for rings in the melt~\cite{Halverson2011c} and even for chromosomes~\cite{Mirny2011} -- which can be modelled as a concentrated suspension of polymers, where loops form due, e.g., to the action of DNA-binding protein bridges. On the other hand, the prediction offered by the slip-link picture is that loops formed by slip-links sliding along rings should lead to both tight (i.e., double-folded) and loose (i.e., swollen) loops, the former scaling with an exponent $\gamma_t = d \nu - \sigma_4 = 2.2$ (here $d$ is the dimensionality, $\nu=0.588$ the SAW exponent and $\sigma_4=-0.46$ the 4-leg vertex exponent~\cite{Duplantier1989,Metzler2002}) and the latter with an exponent $\gamma_l\simeq -0.2$~\cite{Metzler2002}. 

At first sight, the scaling observed in rings in the melt and and in slip-linked chains appear to be different, hence incompatible. However, we analysed the loops formed by the configurations of rings in the melt and characterised each self-contact based on its ``degree'', i.e. on how many internal loops they contained, or, equivalently, on how many internal arcs they contained in their arc representation). The 0-degree loops are the ones which do not contain any other internal loop whereas the highest degree loop is the full contour itself. Through this characterisation we found that 0-degree loops follow a power law 
\begin{equation}\label{tight}
	P_c(l_{\rm 0-degree}) \sim l^{-2.3}
\end{equation}
remarkably close to the one for tight loops in slip-linked chains (Fig.~\ref{fig:sliplinks}B). Instead, summing the probability of self-contact across loops of any degree resulted in a distribution decreasing with the exponent $\gamma \simeq 1.04$, as reported in the literature (Fig.~\ref{fig:sliplinks}B). 

The double scaling found in Figure~\ref{fig:sliplinks}B suggests that the analogy between slip-links might be appropriate, although confined to 0-degree loops. The coexistence of tight loops (as per the scaling in Eq.~\eqref{tight}) and loose loops (evident from contact maps and snapshots of rings) correlates with the abundance of threadings which is at the origin of the topological glass discussed in Section ``\emph{Polymer Melts and Topological Glasses}''. Therefore a possible view is that rings in a melt behave as confined polymers with slip-links, which lead to the coexistence of tight loops threading loose loops. Globally, this structure is not too dissimilar from a tree-like structure (with only tight double-folded conformations), which explains the exponent $\nu \simeq 1/3$. However, locally loose loops may open up to accommodate threadings which become more and more abundant as $M\to \infty$. While speculative, this framework may prove useful in the future to derive to a consistent statistical physics model of ring polymer melts.

\begin{figure*}[t]
	\centering
	\includegraphics[width=0.9 \textwidth]{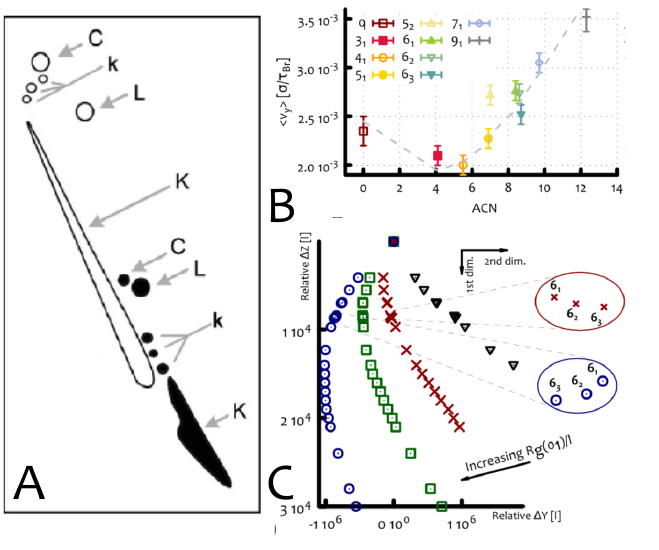}
	\caption{(A) Experimental patterns obtained in 2D gel electrophoresis of viral plasmids in Ref.~\cite{Trigueros2007} (K=knots, C=circular (unknotted), L=linear). (B) Velocity of the knots along the strong field direction, as a function of average crossing number. (B) Simulated 2D gel electrophoresis using a simple biased continuous random walk model~\cite{Michieletto2015pnas}. }
	\label{fig:2dgelep}
\end{figure*}

\subsection{2D Gel Electrophoresis of Knotted DNA}
\label{knots}

One of the most beautiful examples of the way in which topology affects the behaviour of polymers is provided by the electrophoretic mobility of DNA knots~\cite{Stas,Katritch1996a}, and this is the focus of the current Section. Electrophoretic mobility is the principle on the basis of which the knot type of a (nicked, or torsionally relaxed) DNA molecule is determined. The basic idea is that a charged DNA molecule can be dragged through a gel by an electric field, and that its mobility is inversely proportional to the size of the molecule, according to Stokes law. As more complex knots are more compact than simpler knot with the same contour length, it follows that more complex knots will move faster in a gel. More quantitatively, it is well-known that at low electric fields, the electrophoretic speed of knotted bio-molecules scales linearly with the ``average crossing number'' (ACN) of the knot $\mathcal{K}$~\cite{Stas} -- which provides a measure of knot complexity.

Whilst the behaviour at low electric fields is well understood, puzzling results have been obtained for larger electric fields, or in two-dimensional gel electrophoresis~\cite{Arsuaga2005,Trigueros2001,Trigueros2007,Weber2006a,Weber2006b,Cebrian2014}, where the sample is subjected to two different field strength (along two different directions, Fig.~\ref{fig:2dgelep}A). In these cases, it was found that DNA knots display an unexpected non-monotonic mobility as a function of behaviour of their complexities. 

Notably, these surprising results can be reproduced via molecular dynamics simulations of knotted polymers dragged through gels with dangling ends, modelled as imperfect cubic meshes as described in Section~\ref{s_gelentanglement} (i.e. with dangling ends). The presence of dangling ends is relevant to agarose gels in reality, and it is also necessary to reproduce the non-monotonic mobility at intermediate or strong fields (Fig.~\ref{fig:2dgelep}B). 

Our simulations~\cite{Michieletto2015pnas,Michieletto2016phdthesis} showed that as the knotted loops were dragged through this gel, they could get trapped by the dangling ends of the structure (as in the impalament shown Fig.~\ref{fig:piercing}B and discussed in Section~\ref{s_gelentanglement}). 
By performing systematic simulations with different knot types dragged through the gel we measured the average number of times each knot hit the background gel in a given time and for how long their motion was stalled~\cite{Michieletto2015pnas}. We discovered that more complex knots impacted less often with the gel structure, but when they did, they remained stalled for longer time. 
Importantly, by repeating the simulations with different intensities of electric fields we quantified that impalaments occur more frequently for larger values of the electric field~\cite{Michieletto2014ringsoftmatter,Michieletto2015pnas}. 

This behaviour can be explained within the framework of Section~\ref{s_gelentanglement}. On one hand, the average piercing number of knotted rings increases linearly with knot complexity, or ACN (see Fig.~\ref{fig:ent_numb}B). This increases the probability of impalement for complex knots. However, as the ACN increases, the cross section at disposal of the dangling ends to create an entanglement decreases, roughly as the gyration radius, $R_g$ (Fig.~\ref{fig:ent_numb}B). This effect creates a competing trend, as it would lead to a decay in impalement probability with knot complexity. 


We thus argued that knots in gel may be imagined as biased random walkers, with an internal degree of freedom dictated by the corresponding complexity (say, ACN). Higher ACN means that the knot has lower probability to stall (due to an impalement) stalling probability, but larger waiting times once impaled. Because of this trade-off the speed of the knots is not simply linearly increasing with their complexity, but it is in reality a non-monotonic function of ACN (Fig.~\ref{fig:2dgelep}B, the solid line gives the prediction from the random walk theory). In practice, this leads to curved electrophoretic patterns (Fig.~\ref{fig:2dgelep}C), similar to those observed in experiments. 

These results suggest that the shapes of electrophoretic arcs found experimentally, and routinely used to identify knot or link type, or supercoiling degree, in DNA molecules, are profoundly dependent on the DNA-gel interactions. A prediction of this theory is therefore that, if no dangling ends were present, one should no longer observe the non-monotonic mobility -- whilst experiments to probe these are difficult to design, simulations were done and confirmed this prediction~\cite{Michieletto2015pnas}.

\section{Discussions and Conclusions}  

\red{  
Historically, the interest in the rhelogical properties of ring polymer melts have started in the 1980s~\cite{Roovers1985,Klein1986,McKenna1987}. Subsequently, several groups have independently attempted to understand the behaviour of ring tracers embedded in a matrix of linear polymers through concepts of ``threading''. These were intendend as events in which the ring tracer would be pinned, or threaded through by one, or multiple strands of the background polymeric matrix (see Refs.~ \cite{Klein1986,Mills1987,Tead1992,Michieletto2016phdthesis} and therein). The idea that topology could affect the behaviour of polymers in solution ignited the interest of the polymer physics community towards  polymers with non-trivial architectures, such as star, branched, closed, tadpoles, etc.~\cite{Gennes1979,Chremos2015,Antonietti1995,Doi2015a}. In turn, most of these polymeric systems displayed deviations from the standard reptation theory valid for linear chains~\cite{Doi1988}, thereby enhancing the interest towards these systems and highlighting the need for new concepts of ``entanglements'' and ``reptation''. Even several decades after the initial seminal works, the behaviour of ring polymers and its connection to threading topological constraints remains largely obscure and a highly active research field~\cite{Robertson2006,Robertson2007,Yang2010,Chapman2012,Habuchi2013, Shanbhag2017}. 
More recently, a different concept of "threading" have emerged. Now, rings can themselves thread other rings so that the whole system becomes a hierarchically constrained network of topologically constraints. Crucially, this feedback between passive and active threadings is an exquisite feature of melts of rings and cannot be found in systems with different architectures (with the exception of tadpoles). 
}
In light of this, we have covered the latest research aiming to characterise these interactions in melts of rings and to quantify topological entanglements between ring polymers, or between a ring polymer and a disordered environment. We considered a specific set of topological entanglements: inter-ring threadings and impalements between a ring and a rod-like obstacle (modelling, for instance, a dangling end in a physical gel made up by cross-linked polymers, such as agarose). These physical entanglements are difficult to describe mathematically as they can disappear over time, so that normal topological invariants (such as the linking number) cannot be used in a straightforward manner.

Besides describing some of the algorithms to describe threadings and impalements (Fig.~\ref{fig:topint}), we have also discussed physical implications of these architecture-specific entanglements. First, we have considered a concentrated ring polymer solution embedded in a 3D regular mesh (whose unit cells are useful to rigorously define inter-ring threadings). We have seen that as the length of the rings increases (at a fixed value of the overall volume fractions) there is a transition, or crossover, between a regime where threadings are sparse to another one where they percolate through the whole system. Second, we have reviewed computer simulations showing that there is a liquid-glass transition when a sufficiently number of ring polymers in a melt are pinned, or permanently frozen. This kinetic arrest occurs at low volume fraction, much below the ones needed for a glass transition to occur due to excluded volume interactions (jamming) and at temperatures much higher than the ones required to suppress microscopic degrees of freedom (glass temperature). The simulation results we have discussed are also suggestive of a similar transition for a ``standard'' melt of sufficiently long ring polymers, where no rings are pinned. The glassy dynamics is driven by the proliferation of inter-ring threadings (and possibly of self-threadings), hence the corresponding structure is named a ``topological glass''. Third, we have proposed a model according to which a ring polymer in a melt can be seen as an isolated ring polymer with mobile slip-links, where tight loops threads loose ones when the rings become long enough. Finally, we have discussed a potential effect of impalement to the electrophoretic mobility of DNA knots. We have presented evidence here suggesting that the non-monotonic mobility of knots under an intermediate or strong electric field may be due to topological interactions between the rings and the dangling ends of agarose gels. These take longer to resolve for more complex knots, however appear more frequently in the first place for an unknot. The competition between disentanglement time and impalement probability provides a physical mechanism for the non-monotonic mobility.

There are a number of open questions in this fascinating field of topological interactions in ring polymers. First, it would be desirable to generalise the algorithm for threading detection to melts which are not embedded in a regular mesh, or gel. Second, it would be interesting to characterise the self-threadings of long ring polymer in a similar way to the one we have proposed for mutual threadings. Another important open question regards the very existence of the topological glass phase in a melt of ring polymers: it would be of interest to look for this structure either with longer computer simulations than reported here, or in experiments (e.g., with melts of DNA plasmids). 

\appendix

\section{Computational Details}
The simulations discussed in this review are based on the following strategy: We model systems of ring polymers by enclosing $N$ semi-flexible bead-spring polymers formed by $M$ beads in a box with periodic boundary conditions of linear size $L$. The monomer density $\rho = NM/L^3$ is reached by slowly reducing the system size and letting the system equilibrate (each ring needs to travel at least once its own size). The chains are modelled via the Kremer-Grest worm-like chain model~\cite{Kremer1990} as follows: Let $\bm{r}_i$ and $\bm{d}_{i,j}= \bm{r}_j - \bm{r}_i$ be respectively the position of the center of the i-th bead and the vector of length $d_{i,j}$ between beads i and j, the connectivity of the chain is treated within the finitely extensible non-linear elastic model with potential energy,
\begin{equation}
U_{FENE}(i,i+1)=-\dfrac{k}{2}R_0^2\log{\left[ 1-\left( \dfrac{d_{i,i+1}}{R_0}\right)^2\right]}
\end{equation}
for $d_{i,i+1} < R_0$ and $U_{FENE}(i, i + 1) = \infty$, otherwise; here we choose $R_0 = 1.6 \sigma$ and $k = 30 \epsilon/\sigma^2$ and the thermal energy $k_BT$ is set to $\epsilon$. The bending rigidity of the chain is captured with a standard Kratky-Porod potential,
\begin{equation}
U_b(i, i+1, i+2) = \dfrac{k_BTl_p}{2\sigma}\left[ 1 -\dfrac{\bm{d}_{i,i+1} \cdot \bm{d}_{i+1,i+2}}{d_{i,i+1} d_{i+1,i+2}} \right],
\end{equation}
where we set the persistence length $l_p = 5\sigma$. The steric interaction between beads is taken into account by a truncated and shifted Lennard-Jones (WCA) potential
\begin{equation}
U_{LJ}(i,j) = 4\epsilon \left[  \left( \dfrac{\sigma}{d_{i,j}} \right)^{12} - 
\left( \dfrac{\sigma}{d_{i,j}}\right)^6 \right] \theta( 2^{1/6}\sigma - d_{i,j})
\end{equation}
where $\theta(x)$ is the Heaviside function. 
Whenever the background gel is present, it is modelled as a static array of beads following a 3D cubic lattice where half-edges are removed in a random fashion with probability $p$. The beads forming the gel exert purely steric interactions on the beads forming the polymers and do not move as a consequence of thermal noise or kicks received by the polymers. This is a crude (albeit reasonable) approximation for the structure of an agarose gels~\cite{Levene1987} or a gel made of nanowires~\cite{Rahong2014}. Whenever pinned rings are present, these are excluded for the integration, i.e. their beads exert steric interactions but their position is not modified by the force they experience. This crudely models the effect of some external driving force, such as an optical tweezer, which can overcome thermal fluctuations and random collisions within the system~\cite{Gokhale2014}.  
Denoting by $U$ the total potential energy experienced by bead $i$, the dynamic of the beads forming the rings is described by the following Langevin equation:
\begin{equation}
m \dfrac{\partial^2 \bm{r}_i}{\partial t^2} = - \xi \dfrac{\partial \bm{r}_i}{\partial t} - \nabla U + \bm{\eta} \label{eq:langevin}
\end{equation}
where $\xi$ is the friction coefficient of bead $i$ and $\bm{\eta}$ is a stochastic delta-correlated noise. The variance of each Cartesian component of the noise, $\sigma_{\eta}^2$ satisfies the usual fluctuation dissipation relationship $\sigma_{\eta}^2 = 2\xi k_BT$. As customary~\cite{Kremer1990} we set $m/\xi = \tau_{LJ} = \tau_{Br}$, with the LJ time  $\tau_{LJ} = \sigma/m$  and the Brownian time $\tau_{Br} = \sigma^2/D_b$, where
$D_b = k_BT/\xi$ is the diffusion coefficient of a bead of size $\sigma$, is chosen as simulation time step. From the Stokes friction coefficient of spherical beads of diameters $\sigma$ we have: $\xi= 3\pi \eta_{sol} \sigma$ where $\eta_{sol}$ is the solution viscosity which may be useful to map the simulation timescales into real ones, depending on the specific system modelled. The numerical integration of Eq.~\eqref{eq:langevin} is performed by using a standard velocity-Verlet algorithm with time step $\Delta t= 0.01 \tau_{Br}$ and is implemented in the LAMMPS engine~\cite{LAMMPS}.

\textbf{Acknowledgements}
We thank ERC (CoG 648050 THREEDCELLPHYSICS) for support.

\bibliography{EP_knots,MeltGlass,general,papers}

\end{document}